\documentclass[12pt]{article}

\usepackage[english]{babel}
\usepackage{graphicx}
\usepackage{layouts}
\usepackage{siunitx}
\usepackage{mathabx}
\usepackage{lineno}
\usepackage[labelfont=bf]{caption}
\usepackage[left=24mm,top=18mm,right=24mm,bottom=18mm]{geometry}

\let\oldsim\sim 
\renewcommand{\sim}{{\oldsim}}

\def \Ha {H$\alpha$}
\def \MNi {$M_{\text{Ni}}$}
\def \Mej {M_{\rm{ej}}}

\def \td {t_{\rm{d}}}
\def \Msun {M$_\odot$}

\def \lam {$\lambda$}

\def \sn {\mbox{OGLE14-073}}

\newcommand\ion[2]{#1$\;${\scshape{#2}}}

\newenvironment{affiliations}{%
    \setcounter{enumi}{1}%
    \setlength{\parindent}{0in}%
    \slshape\sloppy%
    \begin{list}{\upshape$^{\arabic{enumi}}$}{%
        \usecounter{enumi}%
        \setlength{\leftmargin}{0in}%
        \setlength{\topsep}{0in}%
        \setlength{\labelsep}{0in}%
        \setlength{\labelwidth}{0in}%
        \setlength{\listparindent}{0in}%
        \setlength{\itemsep}{0ex}%
        \setlength{\parsep}{0in}%
        }
    }{\end{list}\par\vspace{12pt}}

\renewenvironment{abstract}{%
    \setlength{\parindent}{0in}%
    \setlength{\parskip}{0in}%
    \bfseries%
    }{\par\vspace{-6pt}}

\begin{document}
\newcommand{\spacing}[1]{\renewcommand{\baselinestretch}{#1}\large\normalsize}
\spacing{2}


\newpage\spacing{1}\setlength{\parskip}{12pt}%
    {\Large\bfseries\noindent\sloppy \textsf{Hydrogen-rich supernovae beyond the neutrino-driven core-collapse paradigm} \par}%

\setlength{\parindent}{0.39in}
\setlength{\parskip}{18pt}
\spacing{2}

\renewcommand*{\thefootnote}{\fnsymbol{footnote}}

{\noindent\sloppy G. Terreran$^{1,2,3}$\footnote{E-mail: gterreran01@qub.ac.uk}~~, 
M.~L. Pumo$^{4,2,5}$,
T.-W. Chen$^{6}$,
T.~J. Moriya$^{7}$,
F. Taddia$^{8}$,
L. Dessart$^{9}$,
L. Zampieri$^{2}$,
S.~J. Smartt$^{1}$,
S. Benetti$^{2}$,
C. Inserra$^{1}$,
E. Cappellaro$^{2}$,
M. Nicholl$^{10}$,
M. Fraser$^{11}$,
{\L}. Wyrzykowski$^{12}$,
A. Udalski$^{12}$,
D.~A. Howell$^{13,14}$,
C. McCully$^{13,14}$,
S. Valenti$^{15}$,
G. Dimitriadis$^{16}$,
K. Maguire$^{1}$,
M. Sullivan$^{16}$,
K.~W. Smith$^{1}$,
O. Yaron$^{17}$,
D.~R. Young$^{1}$,
J.~P. Anderson$^{18}$,
M. Della Valle$^{19,20}$,
N. Elias-Rosa$^{2}$,
A. Gal-Yam$^{17}$,
A. Jerkstrand$^{21}$,
E. Kankare$^{1}$,
A. Pastorello$^{2}$,
J. Sollerman$^{8}$,
M. Turatto$^{2}$, 
Z. Kostrzewa-Rutkowska$^{22,23,12}$,
S. Koz{\l}owski$^{12}$,
P. Mr{\'o}z$^{12}$,
M. Pawlak$^{12}$,
P. Pietrukowicz$^{12}$,
R. Poleski$^{12,24}$,
D. Skowron$^{12}$,
J. Skowron$^{12}$,
I. Soszy{\'n}ski$^{12}$,
M.~K. Szyma{\'n}ski$^{12}$,
K. Ulaczyk$^{12,25}$.
}

\begin{affiliations}
\item Astrophysics Research Centre, School of Mathematics and Physics, Queen's University Belfast, Belfast BT7 1NN, UK 
\item INAF-Osservatorio Astronomico di Padova, Vicolo dell'Osservatorio 5, 35122 Padova, Italy 
\item Dipartimento di Fisica e Astronomia G. Galilei, Universit\`a di Padova, Vicolo dell'Osservatorio 3, 35122 Padova, Italy 
\item Universit\`a degli studi di Catania, Dip. di Fisica e Astronomia, Via Santa Sofia 64, 95123 Catania, Italy 
\item INFN-Laboratori Nazionali del Sud, Via Santa Sofia 62, Catania, 95123, Italy 
\item Max-Planck-Institut f{\"u}r Extraterrestrische Physik, Giessenbachstra\ss e 1, 85748, Garching, Germany 
\item Division of Theoretical Astronomy, National Astronomical Observatory of Japan, National Institutes of Natural Sciences, 2-21-1 Osawa, Mitaka, Tokyo 181-8588, Japan 
\item The Oskar Klein Centre, Department of Astronomy, Stockholm University, AlbaNova, 10691 Stockholm, Sweden 
\item Unidad Mixta Internacional Franco-Chilena de Astronom\'ia (CNRS UMI 3386), Departamento de Astronom\'ia, Universidad de Chile, Camino El Observatorio 1515, Las Condes, Santiago, Chile 
\item Harvard-Smithsonian Center for Astrophysics, 60 Garden Street, Cambridge, MA 02138, USA 
\item School of Physics, O'Brien Centre for Science North, University College Dublin, Belfield, Dublin 4, Ireland 
\item Warsaw University Observatory, Al.~Ujazdowskie~4, 00-478~Warszawa, Poland 
\item Las Cumbres Observatory, 6740 Cortona Drive Suite 102, Goleta, CA 93117, USA 
\item Department of Physics, University of California, Santa Barbara, Broida Hall, Mail Code 9530, Santa Barbara, CA 93106-9530, USA 
\item Department of Physics, University of California, Davis, CA 95616, USA 
\item Department of Physics and Astronomy, University of Southampton, Southampton, SO17 1BJ, UK 
\item Department of Particle Physics and Astrophysics, Weizmann Institute of Science, Rehovot 76100, Israel 
\item European Southern Observatory, Alonso de C\'ordova 3107, Casilla 19, Santiago, Chile 
\item Capodimonte Astronomical Observatory, INAF-Napoli, Salita Moiariello 16, 80131-Napoli 
\item International Center for Relativistic Astrophysics, Piazza delle Repubblica, 10, 65122-Pescara 
\item Max-Planck Institut fur Astrophysik, Karl-Schwarzschild-Str. 1, D-85741 Garching, Germany 
\item SRON, Netherlands Institute for Space Research, Sorbonnelaan 2, 3584 CA Utrecht, the Netherlands 
\item Department of Astrophysics/IMAPP, Radboud University Nijmegen, P.O. Box 9010, 6500 GL Nijmegen, the Netherlands 
\item Department of Astronomy, Ohio State University, 140 W. 18th Ave.,Columbus, OH~43210,~USA 
\item Department of Physics, University of Warwick, Gibbet Hill Road, Coventry, CV4~7AL,~UK 
\end{affiliations}

\makeatletter
\renewcommand{\section}{\@startsection {section}{1}{0pt}%
    {-6pt}{1pt}%
    {\bfseries}%
    }
\makeatother

\renewcommand*{\thefootnote}{\arabic{footnote}}

\clearpage
\begin{abstract}
We present our study of OGLE-2014-SN-073, one of the brightest Type II SN ever discovered, with an unusually broad lightcurve combined with high ejecta velocities. From our hydrodynamical modelling we infer a remarkable ejecta mass of $\mathbf{60^{+42}_{-16}}$~\Msun, and a relatively high explosion energy of $\mathbf{12.4^{+13.0}_{-5.9} \times10^{51}}$~erg. We show that this object belongs, with a very small number of other hydrogen-rich SNe, to an energy regime that is not explained by standard core-collapse (CC) neutrino-driven explosions. We compare the quantities inferred by the hydrodynamical modelling with the expectations of various exploding scenarios, trying to explain the high energy and luminosity released. We find some qualitative similarities with pair-instabilities SNe, although a prompt injection of energy by a magnetar seems also a viable alternative to explain such extreme event.
\end{abstract}

Type II supernovae (SNe) are the final stage of massive stars (above 8~\Msun) which retain part of their hydrogen-rich envelope at the moment of explosion. They typically eject up to $10-15$ \Msun{} of material, with energies of the order of $10^{51}$~erg and peak magnitudes of -17.5~mag \cite{Richardson2002}. Although more luminous events are commonly discovered, their explosion energies are mostly in the range of a few times $10^{51}$~erg, explainable by neutrino-driven explosions and neutron star (NS) formation \cite{Janka2012}. 

OGLE-2014-SN-073 (hereafter \sn{}) is a SN discovered by the Optical Gravitational Lensing Experiment (OGLE-IV) Transient Search\footnote{http://ogle.astrouw.edu.pl/ogle4/transients/} \cite{Wyrzykowski2014a} \cite{Udalski2015} on 2014 August 15.43 UT, at coordinates \mbox{$\alpha_{\rm{J}2000}\,=\,05^{\rm{h}}28^{\rm{m}}51.61^{\rm{s}}$}, \mbox{$\delta_{\rm{J}2000}\,=\,\ang{-62;20;16.05}$}. No stringent constraint on the explosion epoch could be placed, with the last non-detection at $\sim110$~d before discovery. A classification spectrum, taken on 2014 September 24.28 UT \cite{Blagorodnova2014} by the Public ESO Spectroscopic Survey of Transient Objects (PESSTO\footnote{www.pessto.org}) \cite{Smartt2015}, showed very prominent hydrogen P-Cygni features, and no signs of interaction of the ejecta with circumstellar medium. Despite being taken $\sim40$~d after discovery, the temperature and velocities inferred from the spectrum best-matched a Type II SN at $\sim15$~d after explosion, posing a problem on the determination of the actual age of the event.

\section{The host galaxy}
\begin{figure}
\centering
\includegraphics[width=\textwidth]{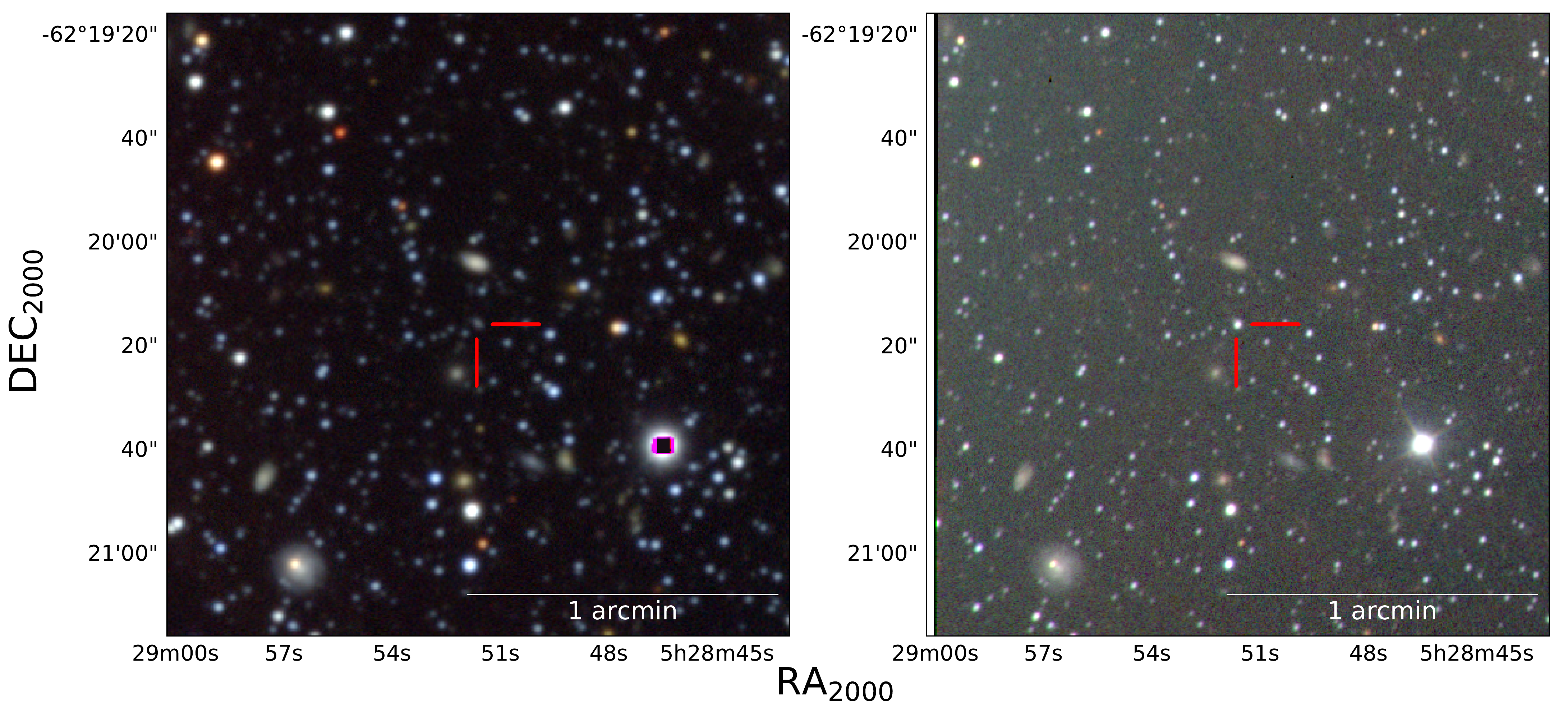}
\caption{\textbf{RGB images of the \sn{} field} $-$ \textbf{(a)}: pre-explosion image taken on 2012 December 22.33 UT by DES during Science verification. SDSS \textit{gri} filters have been used. At the position of \sn{}, marked in red, the faint anonymous host galaxy is observed. \textbf{(b)}: Post-explosion image taken on 2014 September 24.29 UT by PESSTO with NTT+EFOSC2. Johnson-Cousins \textit{BVR} have been used.}
\label{fig: rgb}
\end{figure}

Although \sn{} looked apparently hostless, a pre-discovery image taken on 2012 December 22.33 UT by the Dark Energy Survey (DES) \cite{DES2016} during Science Verification\footnote{http://des.ncsa.illinois.edu/releases/sva1D} showed a faint galaxy at the position of the SN (see Figure \ref{fig: rgb}, left panel). Magnitudes of the host were measured on the available \textit{ugrizy} images, using aperture photometry within \texttt{daophot}. We inferred $g = 23.04 \pm 0.10$~mag, $r = 21.81 \pm 0.16$~mag, $i = 21.98 \pm 0.13$~mag and $z = 21.36 \pm 0.23$~mag. From this observed photometry, we estimated the stellar mass of the host galaxy. We used the stellar population model program \texttt{magphys} \cite{daCunha2008}, which provided a stellar mass of $\log M = 8.7$ M$_{\odot}$, and a $1\sigma$ range from 8.5 to 8.9 M$_{\odot}$ for the host of \sn{}. This is a few times larger than that of the typical mass of the host galaxies of SLSNe with slowly-fading lightcurves \cite{Chen2015}. Following the mass-metallicity relation, this implies a moderately sub-solar metallicity for the host of \sn{}.

A strong contamination from the host galaxy is clearly visible in our last spectrum (see Figure \ref{fig: sp_all}, top panel). From these narrow emissions we could measure a redshift of $z=0.1225$, and from the ratio between \Ha{} and [\ion{N}{ii}] lines \cite{Pettini2004}, we inferred an oxygen abundance of $\text{12+log(O/H)} = 8.36\pm0.10$ for the host galaxy of \sn{}, which is half of the solar-abundance. This estimate, together with the stellar mass previously inferred, are in good agreement with the mass-metallicity relation \cite{Kewley2008}. 

\section{The spectrophotometric evolution}
At a measured redshift of $z=0.1225$, \sn{} peaked at $-19$~mag in the I-band. Very few non-interacting Type II SNe have a luminosity comparable to \sn{}. The $\sim3$ months rise to maximum shown by \sn{} (see Figure \ref{fig: bol_IIP} and SI \S{} 1) and the broad peak of the lightcurve resemble the peculiar Type II SN 1987A \cite{Hamuy1990}, which however was much fainter. After a steep post-maximum decline, the lightcurve of \sn{} settles onto a tail consistent with the decay rate of $^{56}$Co. From the luminosity of this tail, the amount of $^{56}$Ni synthesised during the explosion can be inferred. However, this estimate requires an assumption on the explosion epoch, which is not well constrained. Yet, if we consider that the explosion occurred only the day before discovery, we can derive a solid lower limit \MNi$\geq0.47\pm0.02$~\Msun{}, which is the largest \MNi{} ever estimated for an hydrogen-rich SN \cite{Hamuy2003}. Overall, the spectroscopic evolution of \sn{} (see Figure \ref{fig: sp_all}, top panel) is much slower compared to other Type II SNe (see Figure \ref{fig: sp_all}, bottom panel), with almost no evolution during the $\sim160$~d of spectroscopic follow-up. The spectra are dominated by hydrogen and iron-group elements throughout the entire spectral sequence. Weak forbidden lines start to appear only in the last spectrum, 115~d after maximum. Despite the slow spectroscopic evolution, a progressive cooling of the temperature is visible, as well as a redward shift of the minima of the main absorption features (see SI \S{} 2).

\section{Lightcurve modelling}
In order to investigate the nature of \sn{}, we use the well-tested modelling procedure described in \cite{Pumo2017} and already applied to several other Type II SNe (see Methods for a detailed description). First, an exploratory analysis is conducted in order to determine the parameter space. This is done using the semi-analytical code developed by \cite{Zampieri2003}. The outcomes from this preliminary examination set the framework for the more sophisticated hydrodynamical modelling, that is the general-relativistic, radiation-hydrodynamics Lagrangian code presented in \cite{Pumo2011}. The best fit is obtained by simultaneously comparing (with a $\chi^2$) the bolometric lightcurve, the photospheric gas velocity and continuum temperature of \sn{} with the corresponding quantities simulated by the code.
The resulting best model (shown in Figure \ref{fig: h_mod}) has an explosion energy $E=12.4^{+13.0}_{-5.9} \times10^{51}$~erg, an ejected mass $\Mej=60^{+42}_{-16} $~M$_\odot$ and a radius at explosion $R_0=3.8^{+0.8}_{-1.0} \times10^{13}$~cm (1$\sigma$ confidence level). In the standard CC paradigm, the energy of the explosion results from the neutrino deposition after NS formation \cite{Mueller2016}. Given the low cross-section of the neutrino-matter interaction, these are assumed to deposit only $\sim1$\% of their energy in the ejecta, leading to a fairly robust energy upper limit of $E\lesssim 2\times10^{51}$~erg \cite{Janka2012}. Therefore, in order to achieve the $E\gtrsim 10^{52}$~erg inferred for \sn{} in the context of the neutrino-driven explosions, one has to invoke a much higher, and possibly unphysical neutrino deposition fraction. In addition, the $\Mej$ inferred is several times higher than typical values for Type II SNe \cite{Hamuy2003} \cite{Nadyozhin2003}. We stress that the models are calculated assuming that the SN exploded the day before discovery, and therefore the inferred parameters are all to be considered lower-limits, as all energy, ejecta mass and Ni mass grow moving the explosion epoch back in time. The extraordinary energetics of \sn{}, together with its high $\Mej$ and \MNi{}, are hard to reconcile with the conventional CC scenario.

\section{The pair-instability scenario}
If the progenitor of \sn{} was a very massive star (with He-core between 64 and 133~\Msun{} \cite{Heger2002}), then it could have ended its life due to the instabilities induced by $e^+e^-$ pairs production, in a pair-instability SN (PISN). These events are characterised by very bright (up to $10^{44}$~erg~s$^{-1}$) and broad lightcurves, with rise-times $\gtrsim150$~d, due to the large ejecta masses and hence very long diffusion times \cite{Dessart2013} \cite{Kozyreva2014}. In Figure \ref{fig: bol_pi}, we compare the lightcurve of \sn{} with those of hydrogen-rich PISN models from \cite{Dessart2013}. In particular, we consider a progenitor with zero-age main-sequence (ZAMS) mass $M_{\rm{ZAMS}}=190$~\Msun{} (He-core of $\sim100$~\Msun{}), since, among the models of \cite{Dessart2013}, they produced the dimmest lightcurves (still brighter than \sn{}). Both lightcurves coming from a red supergiant (RSG) and a blue supergiant (BSG) progenitor are considered. The RSG model is brighter at early phases, and lacks the observed rise-time of \sn{}. The decline phase has a very similar slope to that of \sn{}. The BSG progenitor lightcurve shows a reasonable qualitative match, however we lack data before $-100$\,days to probe the full rise (and early peak). \sn{} experiences a faster decline over the first 80~d after peak, but in the tail phase shows similar decline rates, as the RSG model. The tail phase luminosities (Figure \ref{fig: bol_pi}) indicates that the $^{56}$Ni mass in the two PISNe models is much higher than in \sn{}. Our initial estimate of $M_\text{Ni}\geq0.47$~\Msun{} is below the values of 2.6-3~\Msun{} of the models. However we lack a constraint on the explosion epoch of \sn{}, and if we assume that the explosion occurred $\sim90$~d before the initial discovery, \MNi{} could be as high as $\sim1.1$~\Msun{}. While this is still low, it is in the regime of PISN events arising from less-massive progenitors (He-cores $\lesssim90$~\Msun; \cite{Heger2002}). The pre-maximum spectra of the PISN models of \cite{Dessart2013} show many similarities with \sn{}, being dominated by the Balmer lines. However the models show the hydrogen to disappear after the peak. This occurs because at the time of explosion the progenitors of PISNe have very massive He-cores, which prevent the centrally distributed $^{56}$Ni from being mixed to the outer ejecta, where the hydrogen is mainly situated. Instead, in \sn{} the hydrogen dominates the spectrum at all epochs. Therefore, despite the similarities with the models (see also Supplementary Figure 4 for a comparison of temperatures and velocities), late-time spectra of \sn{} are in conflict with a PISN interpretation (unless a source other than $^{56}$Ni is ionising the hydrogen).

Alternatively, in progenitors with smaller He-cores ($\sim30-60$~\Msun{}) than those of PISNe, the instabilities arising from the pairs creation could be insufficient to disrupt the entire star, but violent enough to expel part of the envelope \cite{Woosley2016}. The interaction due to the collision between two (or more) of these shells of material could be an efficient way to power luminous lightcurves, in a so-called pulsational PISN (PPISN; \cite{Woosley2016}). If the shells are dense and massive enough, a photosphere can be created, that could mimic a normal SNe, without clear signs of interaction from the spectra. Given the broad lightcurve of \sn{} and the hydrogen lines visible at all epochs, a scenario with a fast, low-mass inner shell interacting with a slower massive outer one (e.g., see SN 1994W; \cite{Dessart2016}) could perhaps reproduce the observables. Assuming the light curve rise-time to be the diffusion time in the shell, an opacity $\kappa=0.34$~cm$^2$~~g$^{-1}$, an outer radius $R\sim10^{16}$~cm and assuming a constant density $\rho$, using $\td=\kappa\rho R^2/c$ \cite{Arnett1979} we get $M_{\text{shell}}\simeq14$~\Msun{}. Therefore, the outer shell should have a kinetic energy of the order of $10^{52}$~erg. Similar energies can indeed be produced in the most extreme PPISNe \cite{Woosley2016}. However in most models such energies are achieved through large masses and relatively low velocities ($\sim1000-2000$~km~s$^{-1}$), while the first spectrum of \sn{} shows the minimum of the absorption of \Ha{} at $\sim10000$~km s$^{-1}$, probably too high for a pulsation event due to pair-instabilities. Note, in addition, that in this scenario the progenitor star might still be alive, and there would be no $^{56}$Ni synthesised, thus the tail phase match with the radioactive decay of $^{56}$Co would be coincidental. 

\section{The hypernova scenario}
\begin{figure}
\centering
\includegraphics[width=\textwidth]{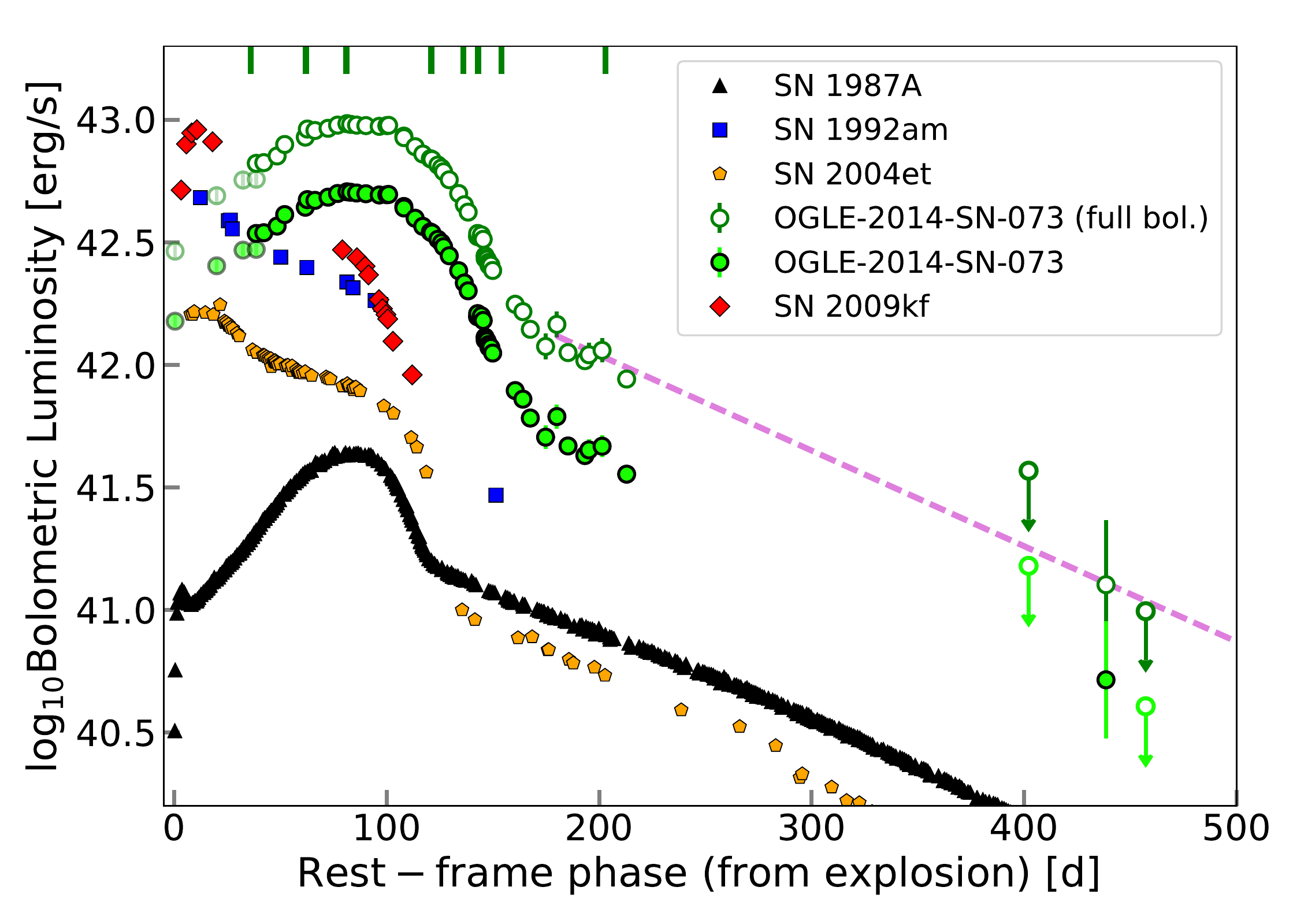}
\caption{\textbf{Bolometric lightcurve of \sn{} and comparison with other Type-II SNe} $-$ Comparison of the optical pseudo-bolometric lightcurve of \sn{} with other luminous non-interacting Type II SNe (Type II-P SNe 1992am, 2004et and 2009kf, and also the peculiar SN 1987A; references in SI \S{} 1). The phase is in rest frame and from explosion, for all SNe apart from \sn{}, for which the first detection is used. The initial 4 points of \sn{} (shaded in the figure) are calculated from only one $I$-band image per epoch, assuming the same SED as for the first epoch with multi-band information. For comparison we include also the full-bolometric lightcurve of \sn{} (see Methods), marked with green hollow circles. The dashed magenta line marks the slope of the $^{56}$Co decay. The green lines at the top of the frame mark the epochs at which the spectra were taken.}
\label{fig: bol_IIP}
\end{figure}

We may notice that, although rare, SNe with $E>10^{52}$~erg do exist. Historically, they have been labelled as ``hypernovae'', and some of them are associated with long gamma-ray bursts (GRBs), e.g. SN 1998bw \cite{Galama1998}. Moreover a hypernova-like explosion has also been invoked to explain the luminous Type II-P SN 2009kf \cite{Botticella2010} \cite{Utrobin2010}. In that case the following parameters were inferred: $\Mej=28$~\Msun{}, $E=22\times10^{51}$~erg and \MNi$=0.40$~\Msun{}. These values are not far from those found for \sn{} (although inferred with a fairly different methodology), and the spectra also show similarities \cite{Botticella2010}. However the lightcurves are quite different (see Figure \ref{fig: bol_IIP}), with that of SN 2009kf resembling more normal Type II-P SNe. In order to try to associate such energetic events within a known scenario, we build a sample of normal Type II SNe, long-rising 1987A-like SNe, standard Ibc SNe (stripped-envelope) and hypernovae, for which an estimate of $E$ and $\Mej$ was available \cite{Taddia2016} \cite{Nadyozhin2003} \cite{Berger2011}, and we plotted these parameters in the top panel of Figure \ref{fig: E_M} (we point out that given the different sources, the methods applied to infer the parameters are quite heterogeneous). The transients appear to gather in 4 clusters, and in particular \sn{} sits in a region characterised by both high $E$ and high $\Mej$, together with SN 2009kf and also with two long-rising SNe, 2004ek and 2004em. This domain of the plot is not populated by ``traditional'' transients. Indeed the ejecta of these 4 SNe are much more massive than that of the hypernovae and are much more energetic than canonical Type II events. Such clustering disappears when comparing $E$ with \MNi{} (Figure \ref{fig: E_M}, bottom panel). Here there seems to be a continuum, with the \MNi{} increasing with $E$, a trend already reported in previous works \cite{Fraser2011} \cite{Kushnir2015}. \sn{} follows the general tendency, however it sits far from all other Type II SNe (with the exception of SN 2009kf), in a region populated by hypernovae. Given the scarcity of the sample, we cannot exclude that high energy Type II SNe may somehow extend towards lower energies or masses, implying, however, that this would either require a much more efficient core-collapse mechanism, or much more massive hydrogen-rich progenitors for hypernovae.

In order to explain the high energy properties of hypernovae, an additional source powering the explosion is required. Such source is usually identified in an ``inner engine'', in the form of a magnetar (e.g. \cite{Thompson2004}) or a black hole (BH; e.g. \cite{MacFadyen2001}). In the first case, a proto NS born with a spin period of the order of 1~ms and with a magnetic field of the order of $10^{15}$~G can inject $10^{52}$~ergs of energy in the inner ejecta in a time scale of $10-100$~s. This energetic shock soon reaches the slower supernova shock, while still travelling through the envelope of the progenitor star, boosting it and producing an hyperenergetic supernova explosion \cite{Thompson2004}. Such an energetic shock has also a deep influence on the nucleosynthesis, as a nickel excess is also expected. Note that this magnetar-engine is different to that supposed to sustain the lightcurves of superluminous SNe \cite{Gal-Yam2012} \cite{Inserra2013}, as in those cases the magnetic field of the NS is one order of magnitude lower, injecting energies of $\sim10^{51}$~erg on a timescale of days to weeks \cite{Kasen2010}. We can speculate that a magnetar with $\rm{B}\geq10^{15}$~G and spin period of $\sim1$~ms could be hidden at the center of the explosion of \sn{}, and this could be the source of energy of the most powerful hydrogen-rich SNe shown in Figure \ref{fig: E_M}. Alternatively, the inner engine can also be constituted by a rapidly rotating BH which, as consequence of the accretion of the matter in-falling from the collapsing progenitor, launches relativistic jets, triggering the explosion.

\section{Conclusions}
Regardless of the energy injection mechanism, the shape of the lightcurve and the spectra of \sn{} unequivocally point to the presence of a massive hydrogen envelope, which is also confirmed by our hydrodynamical modelling. Despite the uncertainties on a definitive determination of the explosion scenario, it appears certain that the progenitor was much more massive than the typical progenitors of Type II SNe \cite{Smartt2009}. However, explosion energies of the order of $10^{52}$~erg and ejecta masses above 50~\Msun{} are too high for a canonical CCSN and neutrino driven explosion. Although there are few other high-energetic hydrogen-rich events which seem to defy the standard CC scenario, \sn{} appears to have an unmatched spectrophotometric evolution. It is puzzling how the progenitor managed to retain such a big amount of its outer envelope, without triggering mass-loss events and transitioning to a Luminous Blue Variable or a Wolf-Rayet star \cite{Meynet2011}. Indeed, progenitors in the mass range that we infer from the ejecta mass should explode as hydrogen-free SNe, according to the current state-of-the-art models. Perhaps a low metallicity environment, like our host galaxy analysis suggested, could have suppressed the mass-loss \cite{Vink2001}. In this context, PISN are supposed to come from massive population III progenitors, however both the PISN and the PPISN scenarios have inconsistencies with the observables of \sn{}. We argued that a central engine scenarios could in principle provide the energy shown by \sn{}, but it opens other issues, like how and why some stars are able to produce compact objects with ultra-intense magnetic fields while others do not. Moreover, it is not clear how these peculiar NSs (or BHs) interact with massive envelopes, especially if jets form, influencing the geometry of the explosion, the $^{56}$Ni mixing and the radiation transport. All together, we believe that the observables of \sn{} give a strong motivation for the search of other similar objects (possibly with better explosion epoch constraints) and for more detailed modelling.

\newcounter{refcount}
\setcounter{refcount}{\value{enumiv}}
\clearpage

\noindent\textbf{Acknowledgements}~~We gratefully thank Profs. M. Kubiak and G. Pietrzy{\'n}ski, former members of the OGLE team, for their contribution to the collection of the OGLE photometric data over the past years.
G.T., S.B., E.C., N.E.-R., A.P. and M.T. are partially supported by the PRIN-INAF 2014 with the project ``Transient Universe: unveiling new types of stellar explosions with PESSTO''. N.E.R. acknowledges financial support by the MIUR PRIN 2010- 2011, ``The dark Universe and the cosmic evolution of baryons: from current surveys to Euclid''. G.T is also supported by the fellowship for the study of bright Type II supernovae, offered by INAF-OaPD.
SJS acknowledges funding from EU/FP7-ERC Grant agreement [291222] and STFC grants ST/I001123/1 and ST/L000709/1. 
T.-W.C. acknowledges the support through the Sofia Kovalevskaja Award to P. Schady from the Alexander von Humboldt Foundation of Germany. 
T.J.M. is supported by the Grant-in-Aid for Research Activity Start-up of the Japan Society for the Promotion of Science (16H07413).
F.T.. and J.S. gratefully acknowledge the support from the Knut and Alice Wallenberg Foundation.
M.F. acknowledges the support of a Royal Society - Science Foundation Ireland University Research Fellowship.
{\L}.W. was supported by Polish National Science Centre grant OPUS 2015/17/B/ST9/03167.
D.A.H. and C.M. are supported by NSF 1313484.
G.D. and M.S. acknowledge support from EU/FP7-ERC grant No. [615929] and the Weizmann-UK 'Making Connections' program.
A.G.-Y. is supported by the EU/FP7 via ERC grant No. 307260, the Quantum Universe I-Core program by the Israeli Committee for planning and funding and the ISF, and Kimmel and YeS awards.
A.J. acknowledges funding by the European Union's Framework Programme for Research and Innovation Horizon 2020 under Marie Sklodowska-Curie grant agreement No 702538. 
K.M. acknowledges support from the STFC through an Ernest Rutherford Fellowship.
Z.K.-R. acknowledges support from ERC Consolidator Grant 647208.
The OGLE project has received funding from the National Science Centre, Poland, grant MAESTRO 2014/14/A/ST9/00121 to AU.
This work is based on observations collected at the European Organisation for Astronomical Research in the Southern Hemisphere, Chile as part of PESSTO, (the Public ESO Spectroscopic Survey for Transient Objects Survey) ESO program ID 197.D.1075, 191.D-0935, and 188.D-3003, and on observations made with ESO Telescopes at the Paranal Observatory under programme 096.D-0894(A).
GEMINI spectra were obtained under the GS-2015A-Q-56 program (P.I. D. A. Howell).
We are grateful to the {\it Istituto Nazionale di Fisica Nucleare - Laboratori Nazionali del Sud} for the use of computer facilities.
This project used public archival data from the Dark Energy Survey (DES). Funding for the DES Projects has been provided by the DOE and NSF (USA), MISE (Spain), STFC (UK), HEFCE (UK), NCSA (UIUC), KICP (U. Chicago), CCAPP (Ohio State), MIFPA (Texas A\&M), CNPQ, FAPERJ, FINEP (Brazil), MINECO (Spain), DFG (Germany) and the collaborating institutions in the Dark Energy Survey, which are Argonne Lab, UC Santa Cruz, University of Cambridge, CIEMAT-Madrid, University of Chicago, University College London, DES-Brazil Consortium, University of Edinburgh, ETH Z{\"u}rich, Fermilab, University of Illinois, ICE (IEEC-CSIC), IFAE Barcelona, Lawrence Berkeley Lab, LMU M{\"u}nchen and the associated Excellence Cluster Universe, University of Michigan, NOAO, University of Nottingham, Ohio State University, University of Pennsylvania, University of Portsmouth, SLAC National Lab, Stanford University, University of Sussex, and Texas A\&M University.
This paper is also based on observations from the Las Cumbres Observatories: we thank their staff for excellent assistance.
IRAF is distributed by the National Optical Astronomy Observatory, which is operated by
the Association of Universities for Research in Astronomy (AURA) under cooperative agreement with the National Science Foundation.

\noindent\textbf{Author contributions}~~G.T. initiated and coordinated the project, managed the follow-up campaign, carried out photometric and spectroscopic analysis, and wrote the manuscript. M.L.P. provided the hydrodynamical modelling and contributed to the manuscript preparation. T.-W.C. performed all the host-galaxy analysis. T.J.M. proposed and investigated the PISN scenario. F.T. identified the similarities of the target with SN 1987A and suggested the scaling. L.D. highlighted the issues with a PISN interpretation and proposed the colliding shells scenario. L.Z. performed the semi-analytical modelling, as preliminary step to the full hydrodynamical modelling. S.J.S. is the PI of the time used at NTT. Moreover, together with S.B., they are the supervisors of the first author, and helped with the coordination of the project and contributing to the manuscript preparation and editing, including final proofreading. C.I. helped with the magnetar hypothesis. E.C. helped with theoretical interpretations, providing very useful advises during manuscript preparation. M.N. retrieved many PISN models, helping with a thorough comparison. M.F. gave useful critics during the manuscript editing and proofread the manuscript. {\L}.W. was the main interlocutor with the OGLE team, providing all the data. D.A.H. was the PI of the GEMINI time from which we obtained two spectra, reduced by C.M. and S.V.. G.D. obtained NTT observations. K.M., M.S., K.W.S., O.Y., and D.R.Y. are PESSTO builders, so helped coordinating the observations at NTT taking care of all the aspects of the PESSTO campaign. J.A., M.D.V., N.E.-R., A.G.-Y., A.J., E.K., J.S., and M.T. provided useful comments and advices after the writing of the first draft of the manuscript. Z.K.-R., S.K., P. M., M.P., P.P., R.P., D.S., J.S., I.S., M.K.S., A.U. and K.U. are all part of the OGLE team and helped obtaining the big amount of data from this survey.

\noindent\textbf{Competing Interests}~~The authors declare no competing financial interests.

\noindent\textbf{Correspondence}~~Correspondence should be addressed to G. Terreran.~(email: gterreran01@qub.ac.uk).

\noindent Supplementary Informations accompanies the paper on www.nature.com/nature
\clearpage

\newpage

\clearpage

\begin{figure}
\centering
\includegraphics[width=0.9\textwidth]{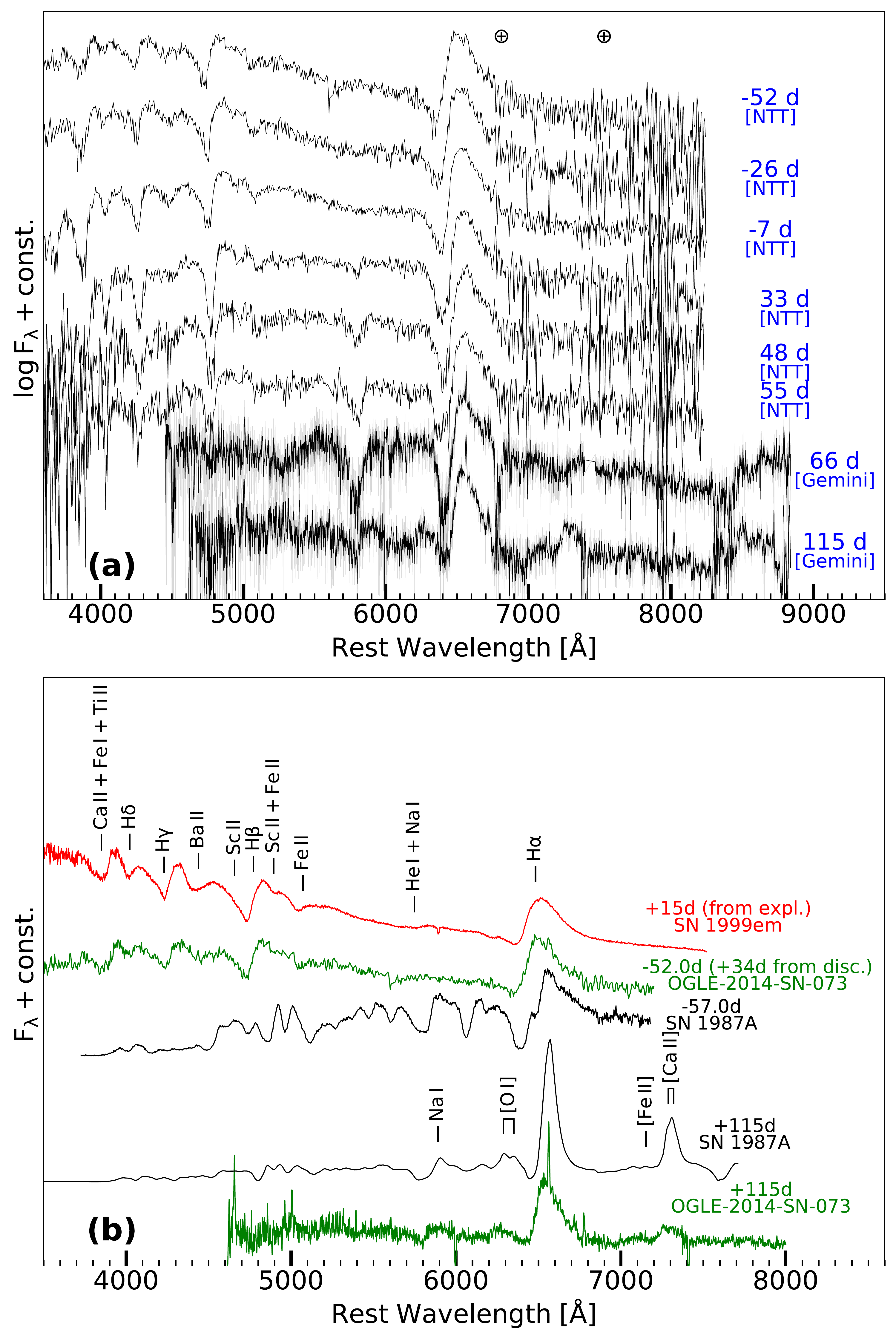}
\end{figure}
\addtocounter{figure}{0}
\begin{figure} 
 \caption{\textbf{Optical spectral evolution of \sn{} and comparison with SN 1987A} $-$ \textbf{(a)}: Optical spectral evolution of \sn{}. The spectra are corrected for reddening and redshift, and shifted vertically for better display. On the right of each spectrum, the phase (in the rest frame) with respect to the bolometric maximum lightcurve and the telescope used are reported. The two GEMINI spectra are smoothed with a boxcar of 5 pixels. The positions of the telluric O$_2$ A and B absorption bands are marked with the $\oplus$ symbol. All spectra will be available on WISeREP (http://wiserep.weizmann.ac.il/home). \textbf{(b)}: Given the similarities between the lightcurves of \sn{} and SN 1987A, we present here the spectroscopical comparison between these two SNe. On the right of each spectrum the phase with respect to the bolometric maximum epoch is reported, unless differently specified. For comparison, also the spectrum of SN 1999em at 15~d after explosion is shown, which was the best match for the classification spectrum. See \mbox{SI \S{} 2} for the references of the objects used for the comparison.}
 \label{fig: sp_all}
\end{figure}
\clearpage

\begin{figure}
\centering
\includegraphics[width=0.9\textwidth]{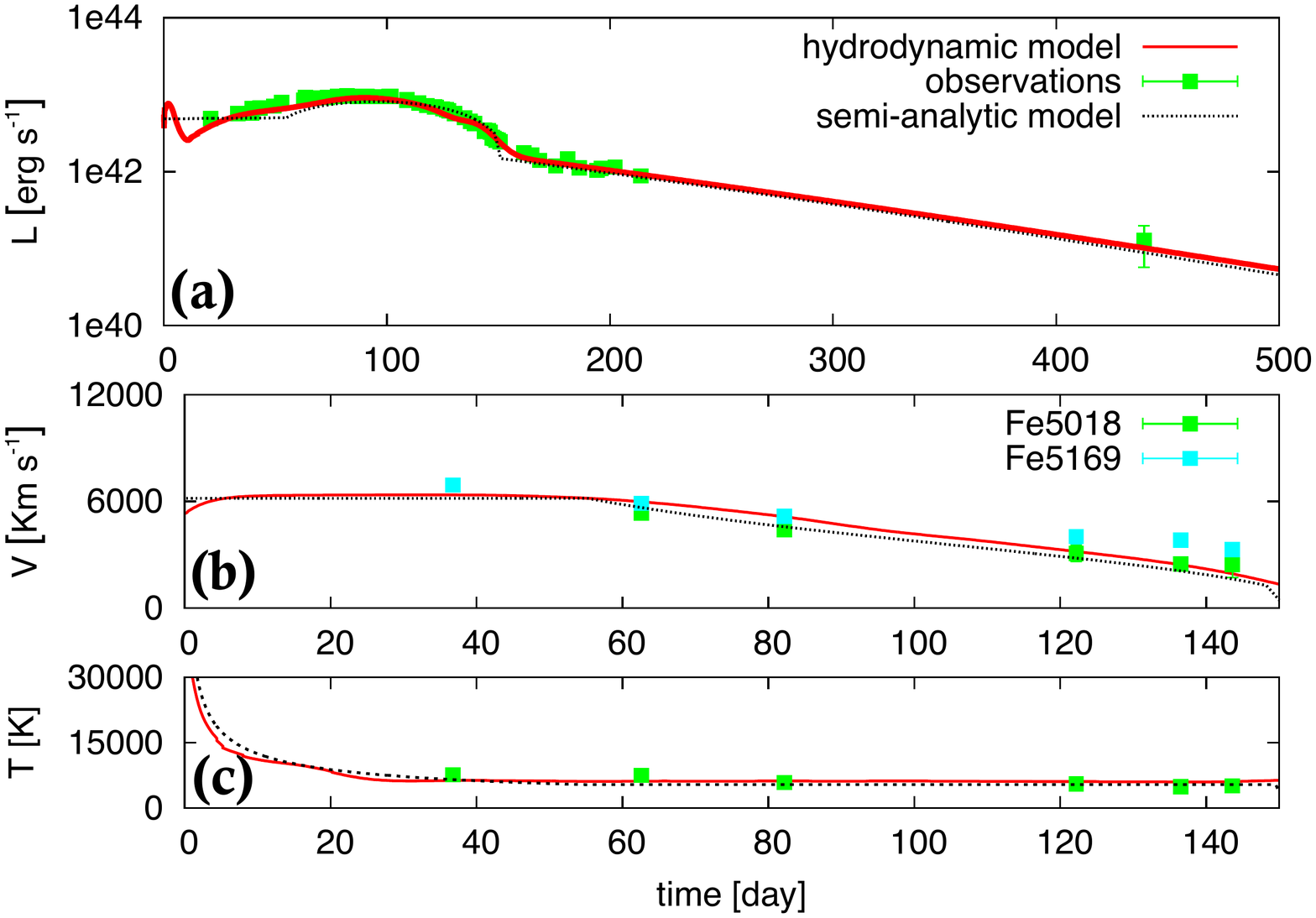}
\caption{\textbf{Hydrodynamical modelling of \sn{}} $-$ Comparison of the evolution of the main observables of \sn{} with the best-fitting model computed with the general-relativistic, radiation-hydrodynamics code described in \cite{Pumo2011}. The best-fitting model parameters are $E= 12.4\times10^{51}$~erg, $\Mej=60$~M$_\odot$ and $R_0=3.8\times10^{13}$~cm. Top, middle and bottom panels show the bolometric light curve, the photospheric velocity and the photospheric temperature evolution, respectively. We assume the explosion to have occurred the day before discovery, and the phase is referred to this epoch. To estimate the photospheric temperature and velocity from observations, we respectively use the continuum temperature and the minima of the profile of the Fe lines, which are considered good tracer of the photospheric 
velocity in Type II SNe. For the sake of completeness, the best-fitting model computed with the semi-analytic code \cite{Zampieri2003} ($E= 21 \times10^{51}$~erg, $\Mej=69$~M$_\odot$ and $R_0=3.5\times10^{13}$~cm) is also shown.}
\label{fig: h_mod}
\end{figure}
\clearpage

\begin{figure}
\centering
\includegraphics[width=\textwidth]{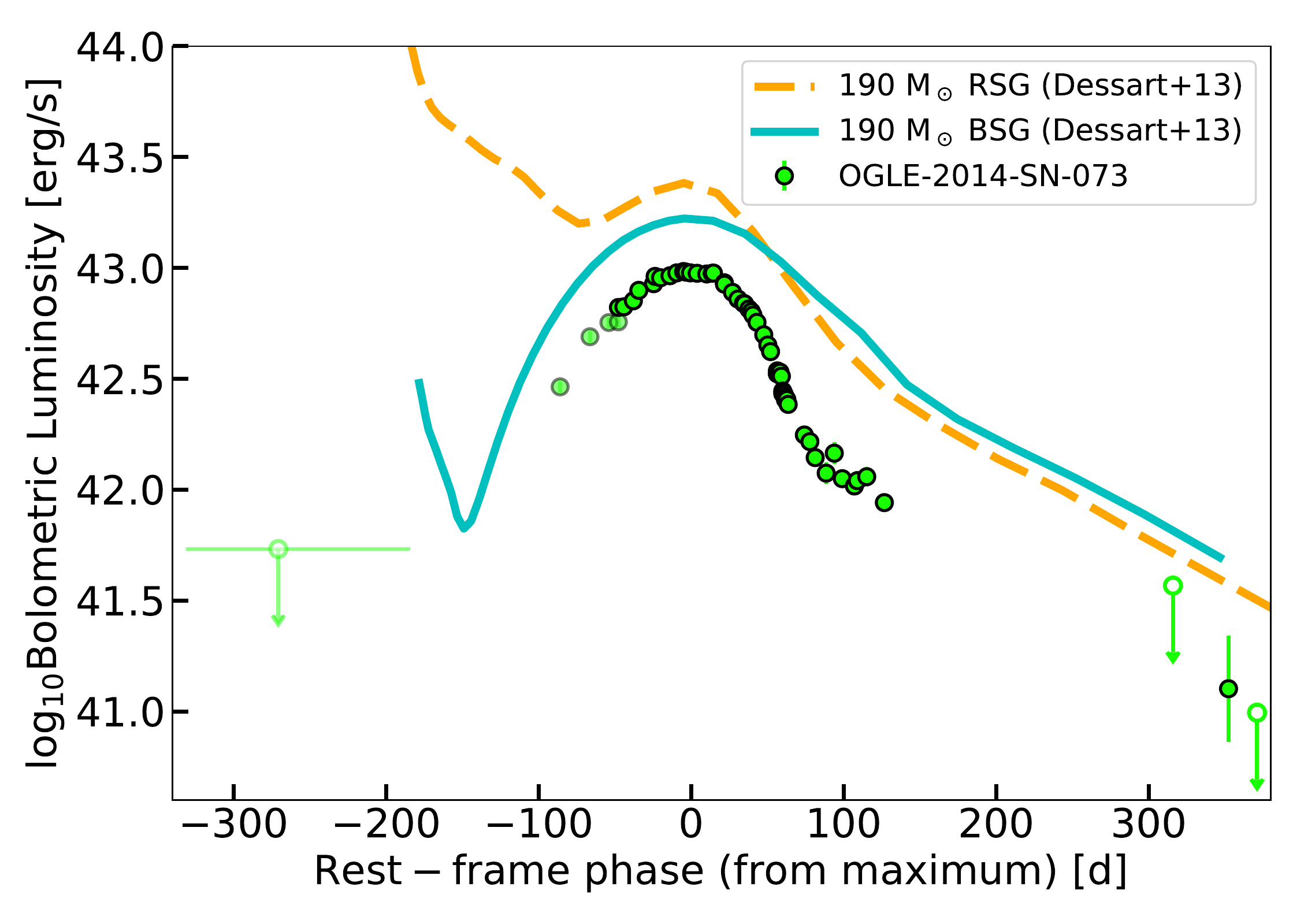}
\caption{\textbf{Comparison of the lightcurve of \sn{} with PISN models} $-$ The green circles mark the bolometric lightcurve of \sn{}. In solid blue and dashed orange, two lightcurve models of PISNe from \cite{Dessart2013}, arising from a BSG and RSG progenitor respectively, both with a $M_{\rm{ZAMS}}=190$~\Msun{}. With the mass-loss prescriptions they used and assuming a metallicity of $10^{-4}$~$\rm{Z}_\odot$, all their models encountered the pair instability when the star was a RSG. The BSG model was produced by artificially truncating the hydrogen envelope just before explosion, in order to simulate stronger mass loss.}
\label{fig: bol_pi}
\end{figure}
\clearpage

\begin{figure}
\centering
\includegraphics[width=0.95\textwidth]{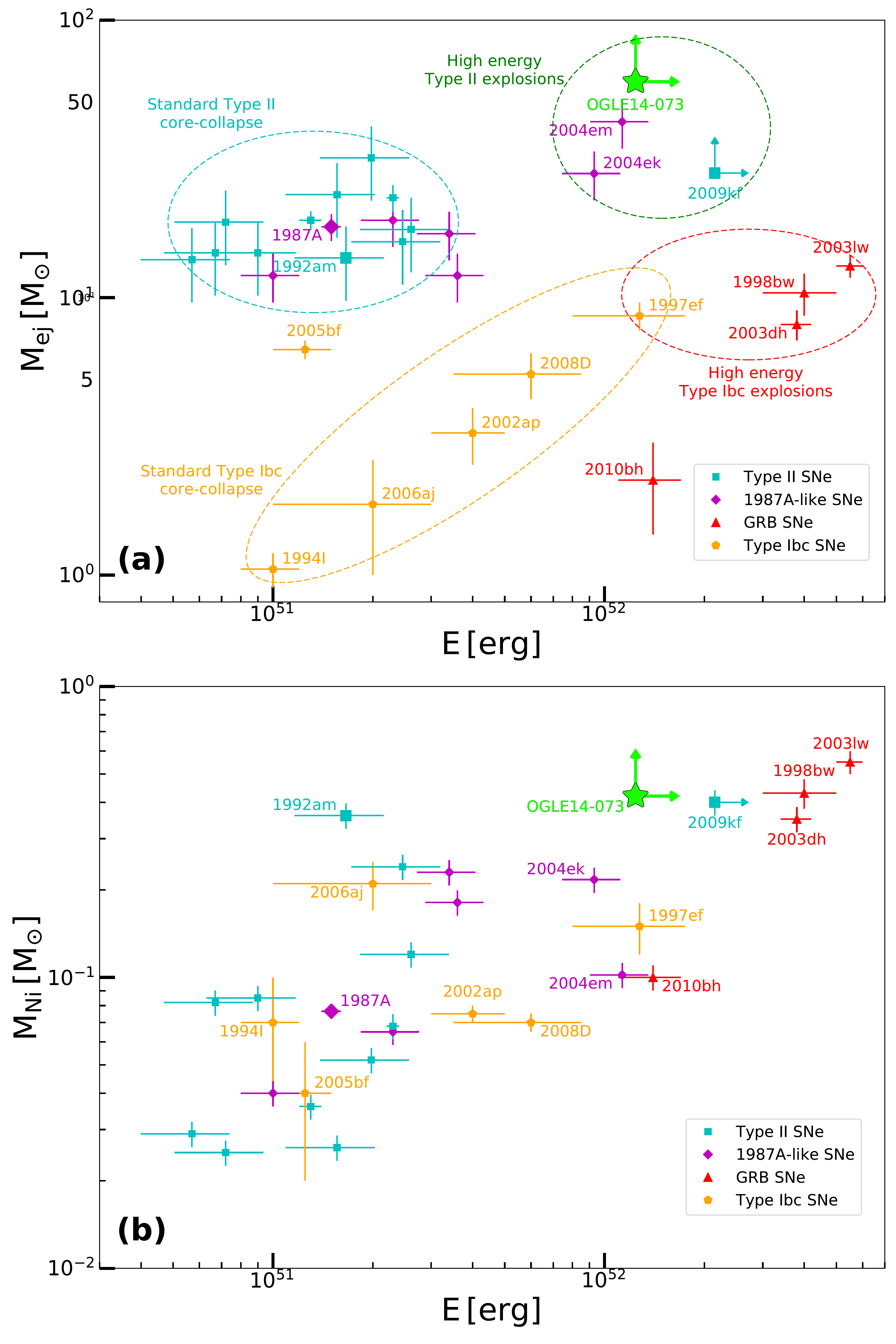}
\end{figure}
\addtocounter{figure}{0}
\begin{figure} 
\caption{\textbf{Explosion energy vs. ejecta mass vs. and vs. $^{56}$Ni mass plots} $-$ \textbf{(a)}: Explosion energy $E$ vs. ejecta mass $\Mej$ plot. A sample of ``Normal'' Type II SNe \cite{Nadyozhin2003}, 1987A-like SNe \cite{Taddia2016}, Type Ibc SNe and SNe related with GRBs (hypernovae; \cite{Berger2011}) are shown. Four major groups are identified in the graph, with \sn{} sitting in a new region of high-energy Type II SNe, together with SNe 2004ek, 2004em and 2009kf. This group of hydrogen-rich SNe appears as a separate cluster from the other standard CCSNe, suggesting a different explosion mechanism for this type of energetic transients. We point out that the sources of the values of $\Mej$ and $E$ are quite heterogeneous, with different methods applied to infer these quantities (see works above referenced).
\textbf{(b)}: Explosion energy $E$ vs. $^{56}$Ni mass \MNi{} plot. A continuum of events is evident, with more energetic events synthesising more $^{56}$Ni. This tendency has already been found in several other works \cite{Fraser2011}, and it led \cite{Kushnir2015} to claim a unique exploding mechanism for all the classes of SNe considered, invoking the collapse-induced thermonuclear explosions (CITE) as an alternative to neutrino-driven explosions \cite{Kushnir2015}. \sn{} seems to respect the general trend, but it sits far from other hydrogen-rich events (with the exception of SN 2009kf).}
\label{fig: E_M}
\end{figure}
\clearpage

\renewcommand*{\thefootnote}{\arabic{footnote}}

\begin{center}
\Large{\underline{Methods}}\\
\end{center}

\section*{Follow-up and data reduction}
OGLE-IV reported the discovery of \sn{} on 2014 September 20.32 UT \cite{Wyrzykowski2014a}, with an $I$-band magnitude of $\sim$19.5~mag. However, inspection of the acquisition images close to the discovery revealed a couple of previous detections, the earliest one being on 2014 August 15.43 UT. We used this date as the discovery reference throughout the paper.
The last non-detection is from OGLE-IV, on 2014 April 27.98 UT (limit 20.0 mag in $I$-band), $\sim110$~d before the first detection. The observational campaign of \sn{} lasted for $\sim$8 months, before it went behind the Sun. Then, when it was visible again, we were able to obtain just one more detection (S/N$\sim4$), in addition to a number of upper-limits. The photometric campaign has been supported also by the acquisition of 8 spectra. A list of the telescopes and instrumentation involved for the follow-up of \sn{} has been reported in Supplementary Table 1.

Images from the Las Cumbres Observatory\footnote{http://lcogt.net/} (LCO; \cite{Brown2013}) and from OGLE-IV were automatically ingested and reduced using the \textsc{lcogtsnpipe} pipeline \cite{Valenti2016} and the OGLE-IV Data Analysis System \cite{Udalski2015} respectively. We reduced all the images coming from the other telescopes by correcting for overscan, bias and flatfields, using standard procedures within \textsc{iraf}\footnote{IRAF is distributed by the National Optical Astronomy Observatory, which is operated by the Association of Universities for Research in Astronomy (AURA) under a cooperative agreement with the National Science Foundation. http://iraf.noao.edu/}. The NIR images, all coming from NTT+SOFI, were reduced using the PESSTO pipeline \cite{Smartt2015}. For the photometric measurements, the \textsc{SNOoPY}\cite{Cappellaro2014} package has been used, which allowed, for each exposure, to extract the magnitude of the SN with the point-spread-function (PSF) fitting technique, using \textsc{daophot} \cite{Stetson1987}. If the transient was not detected in the image, conservative upper-limits were estimated, corresponding to a S/N of 2.5. To derive the magnitude of the SN, we first estimated the zero point and the colour term of the night through the observation of photometric standard fields \cite{Landolt1992}. Then we calibrated a sequence of secondary stars in the field of \sn{}, which were subsequently used to calibrate the SN in each night. For NIR images, we used as reference for the calibration the Two Micron All Sky Survey (2MASS) catalog\footnote{http://www.ipac.caltech.edu/2mass/} \cite{Skrutskie2006}. 
Finally, we applied a $K$-correction computed from the sequence of spectra we gathered. Error estimates were obtained through an artificial star experiment, combined (in quadrature) with the PSF fit error returned by \textsc{daophot}, and the propagated errors from the photometric calibration. SDSS $griz$ filters were used in 3 epochs taken at LCO, and we converted the extracted magnitudes to Johnson-Cousins \textit{BVRI} filters, following the relations derived by \cite{Chonis2008}. The $i$ filter of EFOSC2 is actually a Gunn $i$, nevertheless it has been calibrated as a Cousins $I$. All the magnitudes reported in this work are calibrated in the Vega system. OGLE-IV provided a great number of images where the SN was not detectable (both pre-explosion images and images taken after the SN faded below their detection limit). We stacked them in 3 deeper images, one of which showed a detection.

Given the contamination of the galaxy in the last epochs, a template subtraction would be appropriate. However, the pre-discovery images were either not deep enough (the ones from OGLE) or with different filters than those used for the followup (the one from DES). The SN was not detected in the very last epoch taken with VLT+FORS2 (using \textit{VRI} filters), so we decided to use this last acquisition as the template. We performed the subtraction using \textsc{hotpants}\footnote{http://www.astro.washington.edu/users/becker/v2.0/hotpants.html} by PSF matching of the field stars. We note that the epoch used as template was taken only 75~d after the last detection in $R$ and $I$-band. Thus it is likely that the SN flux was not completely negligible yet, resulting in an over-subtraction of the actual SN signal. For this reason, the magnitude measurements at 400~d in the $RI$ bands are to be considered as upper-limits. 

The bolometric and pseudo-bolometric lightcurves of \sn{} were built first converting the broad-band magnitudes into fluxes at the effective filter wavelengths, building in this way the spectral energy distribution (SED). If at some epoch a particular band was missing, its flux was inferred by assuming constant colour or by interpolation from close-by detections. Then we integrated the SED using the trapezoidal rule, assuming zero flux at the integration boundaries. Since our photometry covered mainly the optical wavelengths, in order to create a full bolometric lightcurve we had to apply a bolometric correction. We estimated it by fitting the SED (to estimate the bolometric correction we used only the SED measured from epochs where the SN was detected in more than 2 bands) with a blackbody and then adding the missing flux, measured from 0 to $\infty$, to the optical luminosities. 

For the optical spectra, the extractions were done using standard \textsc{IRAF} routines. The spectra of comparison lamps and of standard stars acquired on the same night and with the same instrumental setting were used for the wavelength and flux calibrations, respectively. A cross-check of the flux calibration with the photometry (if available from the same night) and the removal of the telluric bands with the standard star were also applied. The GEMINI spectra were reduced using a combination of the \textsc{Gemini IRAF} package and custom scripts in Python\footnote{https://github.com/cmccully/lcogtgemini/}. We performed overscan and master bias subtraction and corrected for the quantum efficiency difference between the chips using \textsc{Gemini IRAF} tasks. We removed any remaining differences in the inter-pixel sensitivity from lamp flat field images. Pixels affected by cosmic rays were identified using the \textsc{astroscrappy} package\footnote{https://github.com/astropy/astroscrappy}.

Note that given the distance of \sn, a time-dilation correction has been applied, and all the phases reported are always to be considered in rest-frame, unless explicitly expressed.

The host galaxy analysis has been done on pre-discovery \textit{ugriz} images taken on 2012 December 22.33 UT by DES, during Science Verification. No flux from the SN is assumed to be present at this time. Magnitude measurements of the host were carried out using aperture photometry within {\sc iraf/daophot}. We let the aperture size vary until we were confident that it encompassed the whole host flux and avoided other nearby objects. The aperture radius adopted was $\sim$\ang{;;2}. The zeropoint was determined with 55 reference stars in the field ($\sim$\ang{;3;} around the host) which were also used for the SN photometry calibration. The inferred host apparent magnitudes are $g = 23.04 \pm 0.10$~mag, $r = 21.81 \pm 0.16$~mag, $i = 21.98 \pm 0.13$~mag and $z = 21.36 \pm 0.23$~mag.

After the Milky Way extinction correction ($A_{V} = 0.17$ mag; \cite{Schlafly2011}), we applied the luminosity distance of 573.9 Mpc ($z = 0.1225$; \cite{Wright2006}) using a cosmology of H$_{0} = 70$\,km\,s$^{-1}$\,Mpc$^{-1}$, $\Omega_{\rm M} = 0.27$, $\Omega_{\rm \lambda} = 0.73$ to calculate the host flux. 
We employed the {\sc MAGPHYS} stellar population model program of \cite{daCunha2008} to estimate the stellar mass from the observed photometry of the host galaxy. This code employs a library of stellar evolution and population models from \cite{Bruzual2003} and adopts the Galactic disc initial mass function (IMF) of \cite{Chabrier2003}. {\sc MAGPHYS} first found the best-fit galaxy model ($\chi_{\text{red}}^{2} = 1.4$), and then calculated the probability density function over a range of model values, inferring the median of stellar mass of $10^{8.7}$ M$_{\odot}$, and a $1\sigma$ range from $10^{8.5}$ to $10^{8.9}$ M$_{\odot}$ for the host of \sn{}. This stellar mass is a few times more than host galaxies of some SLSNe with slowly-fading lightcurves (e.g. \cite{Chen2015}). Following the mass-metallicity relation, this implies a sub-solar metallicity for the host of \sn{}. We noticed some flux excess of the observed $r$-band while comparing the best-fit model, which indicate the host may have a strong contribution from [\ion{O}{iii}] lines, as it is also confirmed by our last spectrum of \sn{} (see Figure \ref{fig: sp_all}, top panel).

We used the spectrum at +115d after maximum for measuring the emission line flux from the host galaxy (see top panel of Figure \ref{fig: sp_all}). The contamination from the SN is strong and the H$\beta$ line is not detected. Hence, we used the N2 method \cite{Pettini2004} for the oxygen abundance. Given the close wavelengths of \Ha{} and [\ion{N}{ii}] lines, this has the advantage of being less affected by dust extinction. We inferred an oxygen abundance of $\text{12+log(O/H)} = 8.36\pm0.10$ for the host galaxy of \sn{}, which is equal to 0.5 solar-abundance (assuming a solar abundance of 12+log(O/H) = 8.69; \cite{Asplund2009}). This estimate, together with the stellar mass previously inferred, are in good agreement with the mass-metallicity relation \cite{Kewley2008}. Nevertheless, a future pure, deep host spectrum is required to measure the host metallicity more accurately.

We also measured the star-formation rate (SFR) \cite{Kennicutt1998} of the host galaxy from the \Ha{} luminosity ($2.41\times10^{39}$ erg\,s$^{-1}$) and then divided it by 1.6 assuming a Chabrier initial mass function. The SFR of the host is $> 0.01$~\Msun{} yr$^{-1}$, and the specific SFR (stellar mass/SFR) is $>0.02$~Gyr$^{-1}$. We point out that this is actually a lower limit, since we did not apply any internal dust extinction correction, and the \Ha{} flux is contaminated by the SN flux.

The data that support the plots within this paper and other findings of this study are available from the corresponding author upon reasonable request.

\section*{Modelling procedure}
The ejected mass $\Mej$, the progenitor radius at the explosion $R_0$ and the total (kinetic plus thermal) explosion energy $E$ of \sn{} are estimated through a well-tested modelling procedure, which is thoroughly described in \cite{Pumo2017} and \cite{Zampieri2007}. This procedure includes the hydrodynamical modelling of all the main SN observables (i.e. bolometric light curve, evolution of line velocities and the temperature at the photosphere), where $\Mej$, $R_0$ and $E$ are derived from a simultaneous $\chi^2$ fit of these observables to the model calculations. The full radiation-hydro models are calculated with the code presented in \cite{Pumo2010} and \cite{Pumo2011}. It is able to simulate the evolution of the physical properties of SN ejecta and reproduce the behaviour of the main SN observables, from the breakout of the shock wave at the stellar surface up to the radioactive-decay phase. The radiative transfer is accurately treated at all optical depth regimes, by the coupling of the radiation moment equations with the hydrodynamics equations. A fully implicit Lagrangian finite difference scheme is adopted to solve the energy equations and the radiation moment equations. The description of the ejecta evolution takes into account the heating effects due to the decays of the radioactive isotopes synthesised during the SN explosion. The gravitational effects of the compact remnant are also considered through a fully general-relativistic approach. The initial conditions used in the code well mimic the physical properties of SN progenitor after the shock breakout at the stellar surface and the reverse shock passage through the ejecta, with the exception of the outermost high-velocity shell of the SN ejecta which can recombine quickly and is not included. The latter can provide a non-negligible contribution to the early emission of the SN, preventing to accurately reproduce the evolution of the photospheric velocity at early phases, but it is typically not crucial for the total mass-energy budget of the supernova. Including it would likely increase both the estimated ejected mass and explosion energy, reinforcing the idea that OGLE14-073 is an extraordinary object. In particular, the initial density profile is described by Equation 6 of \cite{Pumo2011}. It is derived from the so-called radiative zero solution of \cite{Arnett1980} (that well approximates the initial temperature profile) assuming that the ejecta are radiation-dominated.

However, the computation of grid of models with the full radiation-hydro code is very time-consuming. Therefore, we need first to constrain the parameter space of the SN progenitor and ejecta. This is accomplished by means of the semi-analytical model described in \cite{Zampieri2003} and \cite{Zampieri2007}, that solves the energy balance equation for ejecta of constant density and free-coasting (in homologous expansion). The plasma is assumed to be dominated by the radiation pressure. Both the recombination of the ionised matter and the decay of the $^{56}$Ni and $^{56}$Co synthesised during the explosion are considered as sources of heating of the ejecta. The parameters estimate is carried out as described above, fitting the main SN observables to model calculations using $\Mej$, $R_0$ and $E$ (or the initial expansion velocity) as fitting parameters. This preliminary analysis yields as the best parameters $E= 21^{+29}_{-13} \times10^{51}$~erg, $\Mej=69^{+52}_{-36}$~M$_\odot$ and $R_0=3.5^{+0.8}_{-1.1}\times10^{13}$, at a confidence level of 3$\sigma$.

Once an approximate but reliable estimate of the physical conditions describing the SN progenitor at explosion is obtained, such reduced framework is used as start for the above-mentioned general-relativistic,radiation-hydrodynamics Lagrangian modelling. The parameters resulting from this modelling are $E= 12.4^{+13.0}_{-5.9}\times10^{51}$~erg, $\Mej=60^{+42}_{-16}$~M$_\odot$ and $R_0=3.8^{+0.8}_{-1.0} \times10^{13}$~cm (1$\sigma$ confidence level). The reported uncertainties are an estimate of the errors related to the $\chi^2$ fitting procedure used for the modelling, and are inferred following the same approach as described in \cite{Pumo2017} but considering $1\sigma$ confidence intervals. Usually the typical values of these uncertainties are in the range $\sim10-30$ per cent (relative error) for conventional Type II SNe. However, the unique characteristics of \sn{} inflated these uncertainties, providing wide bounds for the inferred parameters, nevertheless reflecting a reliable and solid range of values. We note in particular that the upper error for the energy is of the order of 100\%, suggesting that is more probable for the explosion to be more energetic with respect to what we inferred, rather than less energetic. We also point out that the inferred errors do not include possible systematic uncertainties linked to the input physics (e.g. opacity treatment, approximate initial condition of our models) nor uncertainties on the assumptions made in evaluating the modelled observational quantities (e.g. the adopted reddening, explosion epoch and distance modulus). Although the variations of the parameters $E$, $\Mej$ and $R_0$ due to
these systematic uncertainties may be not negligible, they do not have a significant impact on the overall results (see also Sect. 2.1 of \cite{Pumo2017} and references therein for further details). In the case of \sn{} they can produce a systematic increase of E (and $\Mej$) reinforcing the idea that \sn{} is an extraordinary object which defies the canonical neutrino-driven core-collapse paradigm.

The parameters inferred from the semi-analytical model and the more accurate hydrodynamical modelling are in good agreement. The explosion energy is approximately 70\% higher in the semi-analytical mode because the latter does not take into account the ejecta acceleration that occurs in the first few days after explosion and that converts most of their internal energy into kinetic energy. To correctly reproduced the velocity profile, the semi-analytical code requires a larger initial velocity, hence leading to an overestimate of the initial kinetic and total explosion energy.

\setcounter{refcount}{\value{enumiv}}

\clearpage

\renewcommand{\thepage}{S-\arabic{page}}
\setcounter{page}{1}

{\Large\bfseries\noindent\sloppy \textsf{Hydrogen-rich supernovae beyond the neutrino-driven core-collapse paradigm} \par}
\vspace{1cm}
\begin{center}
\Large{\underline{Supplementary Information}}\\
\end{center}

\clearpage

\renewcommand{\figurename}{Supplementary Figure}
\renewcommand{\tablename}{Supplementary Table}
\setcounter{figure}{0}
\setcounter{table}{0}

\section*{S1~~~Photometry}
The photometric evolution of \sn{} in all bands is shown in Supplementary Figure \ref{fig: snlc}. The explosion epoch is not constrained up until $\sim100$~d before the first detection, and the limiting magnitude of the last non-detection is not particularly stringent either. We obtained a deeper image by stacking together $\sim5$~months of pre-explosion images, inferring a lower-limit of $>-18.9$~mag in $I$-band. Since discovery, all bands show a very slow rise to maximum, which is reached after $>86$~d. In the $I$-band, which is the best covered band, the lightcurve flattens for $\sim50$~d and then drops by $\sim1.5$ ~mag in about 20~d, before settling on a less steep tail. We note that the drop in magnitude is much deeper in $V$ and $R$-band, with no information for the $B$-band as we do not have a detection on the tail. The drop in magnitude is also visible in the NIR, however with only 3 epochs.

In Figure \ref{fig: bol_IIP}, we show the bolometric curve built with the procedure described in Methods, together with the optical-only pseudo-bolometric lightcurve. Fitting the peak with a low order polynomial, we inferred the maximum of the bolometric lightcurve, which occurred on $\rm{MJD}=56982.7\pm1.9$, $\sim86.3$~d in rest-frame after the discovery. Given the lack of information on the explosion epoch, we used this epoch as reference throughout the paper, otherwise explicitly reported.

After $\sim$150~d the lightcurve seems to settle onto the so-called radioactive tail, i.e. when the SN is powered by trapping of $\gamma$-rays and positrons originating from the $^{56}$Co radioactive decay. The luminosity of this tail can give us an indirect measurement of the mass of $^{56}$Ni (parent element of the cobalt) synthesised by the SN (see \cite{Cappellaro1997}). However, since this estimate has a time-dependence linked to the e-folding time of the radioactive elements involved, the measurement is also dependent on the explosion epoch. Assuming the explosion occurred just the day before discovery we then estimate a $^{56}$Ni mass of \MNi$\geq0.47\pm0.02$. 

In Figure \ref{fig: bol_IIP} we present a comparison of the optical pseudo-bolometric lightcurve of \sn{} with those of the other bright Type II-P SNe 1992am \cite{Schmidt1994}, 2004et \cite{Maguire2010} and 2009kf \cite{Botticella2010}, and we include also the peculiar Type II SN 1987A \cite{Hamuy1990}. It strikes us immediately how luminous \sn{} is, with only SN 2009kf outshining it. All the other SNe considered, despite showing above-average luminosities, present quite normal Type II-P lightcurves. However, the behaviour of \sn{} is different, showing a very slow rise to maximum, which is associated more with the peculiar SN 1987A. Indeed the rise-time of these two transients is even comparable. We thus investigate the hypothesis that \sn{} was actually a scaled-up 1987A-like event, and we compared its lightcurve with those of the sample of long-rising Type II SNe from \cite{Taddia2016}. The comparison is shown in Supplementary Figure \ref{fig: bol_87A}. Despite the overall similar shape, both the time scale and the luminosity do not match, as \sn{} presents a much broader and much brighter lightcurve than any other SN in the sample. Therefore the photometric evolution shown by \sn{} is not equalled by any other Type-II SN observed so far.

\section*{S2~~~Spectroscopy}
Figure \ref{fig: sp_all} (top panel) shows the complete optical spectral evolution of \sn{}. The spectra are all relatively red, although this may be due to some line-blanketing on the bluer part. It is evident that there is almost no change for the $\sim150$~d of spectroscopic follow-up. Prominent Balmer lines are present since the first spectrum, along with some faint \ion{Fe}{ii}, probably blended with other metal lines like \ion{Sc}{ii} and \ion{Ba}{ii}. The \ion{Ca}{ii} H\&K lines are also present. The \ion{He}{i}$+$\ion{Na}{I} absorption blend at $\sim5750$~\AA\ starts to appear in the spectrum at $33$~d after maximum, increasing its intensity with time. In the last two spectra, the \ion{Ca}{ii} infrared triplet \lam\lam8498,8542,8662 is clearly visible, however the temporal information on its appearance is not available, as this region is cut out from the previous spectra. Finally, in the very last spectrum the [\ion{Ca}{ii}] \lam\lam7291,7324 starts to appear, and the emission feature on the blue side of \Ha{} is likely a hint of [\ion{O}{i}] rising up, indicating that the SN is approaching the nebular phase. In particular, the ratio between the \ion{Ca}{ii} forbidden doublet and the NIR triplet is $\sim0.6$. This value suggests an electron density of the emitting region $N_e\simeq10^8$~cm$^{-3}$ \cite{Fransson1989}, which would also explain the weak [\ion{O}{i}] $\lambda\lambda$6300,6364 doublet. From the FWHM of the lines, we inferred velocities of $\sim7100$~km s$^{-1}$ for \Ha{}, $\sim4500-4800$~km s$^{-1}$ for the Ca, and $\sim2800$~km s$^{-1}$ for the O. Such stratification points towards an extended hydrogen outer layer. Unfortunately, after this last spectrum, the SN went behind the Sun and when it was visible again it was to too faint take a fully nebular spectrum. In the last spectrum, narrow lines coming from the host galaxy are visible, which we used to measure the redshift $z=0.1225$.

A comparison of 2 spectra of \sn{} at -52 and +115~d with respect to the maximum light (e.g. $34$~d and $201$~d from discovery respectively) with the spectra of SNe 1987A \cite{Catchpole1987} and 1999em \cite{Elmhamdi2003} (best match for the classification spectrum) at similar phases is reported in Figure \ref{fig: sp_all} (lower panel). \sn{} looks much less evolved with respect SN 1987A, at both the two phases considered. At 57~d before maximum, the latter SN shows a forest of metal lines (e.g. \ion{Sc}{ii}, \ion{Ti}{ii}, \ion{Ba}{ii}, \ion{Cr}{ii}) which are not that evident in the spectra of \sn{}, which appear much more metal-poor. At 115~d after maximum, SN 1987A presents well developed forbidden lines, while in \sn{} they are just starting to arise. Moreover, the \Ha{} in the latter SN is larger than the former one, indicating still rapidly expanding ejecta.

From Figure \ref{fig: sp_all} (top panel), a temperature evolution is also evident, as well as an evolution in velocity of the few lines present. We report this evolution in Supplementary Figure \ref{fig: super}, as well as a comparison with SN 1987A and the best sampled long-rising SNe from the sample of \cite{Taddia2016}. The velocities have been measured from the position of the minimum of the absorption of each feature. The temperatures have been estimated with a black-body fit to the continuum of the spectra. Using the peak as reference epoch, the velocities appear to be higher with respect to SN 1987A but compatible with the other two SNe presented. However, the decline looks quite steady, different for example than SN 1987A and SN 2004ek which showed a steeper decline at early times. Also \sn{} seems to be in line with the temperatures of other 1987A-like events. All spectra will be available on WISeREP (http://wiserep.weizmann.ac.il/home) \cite{Yaron2012}. 

\clearpage

\begin{figure}
\centering
\includegraphics[width=\textwidth]{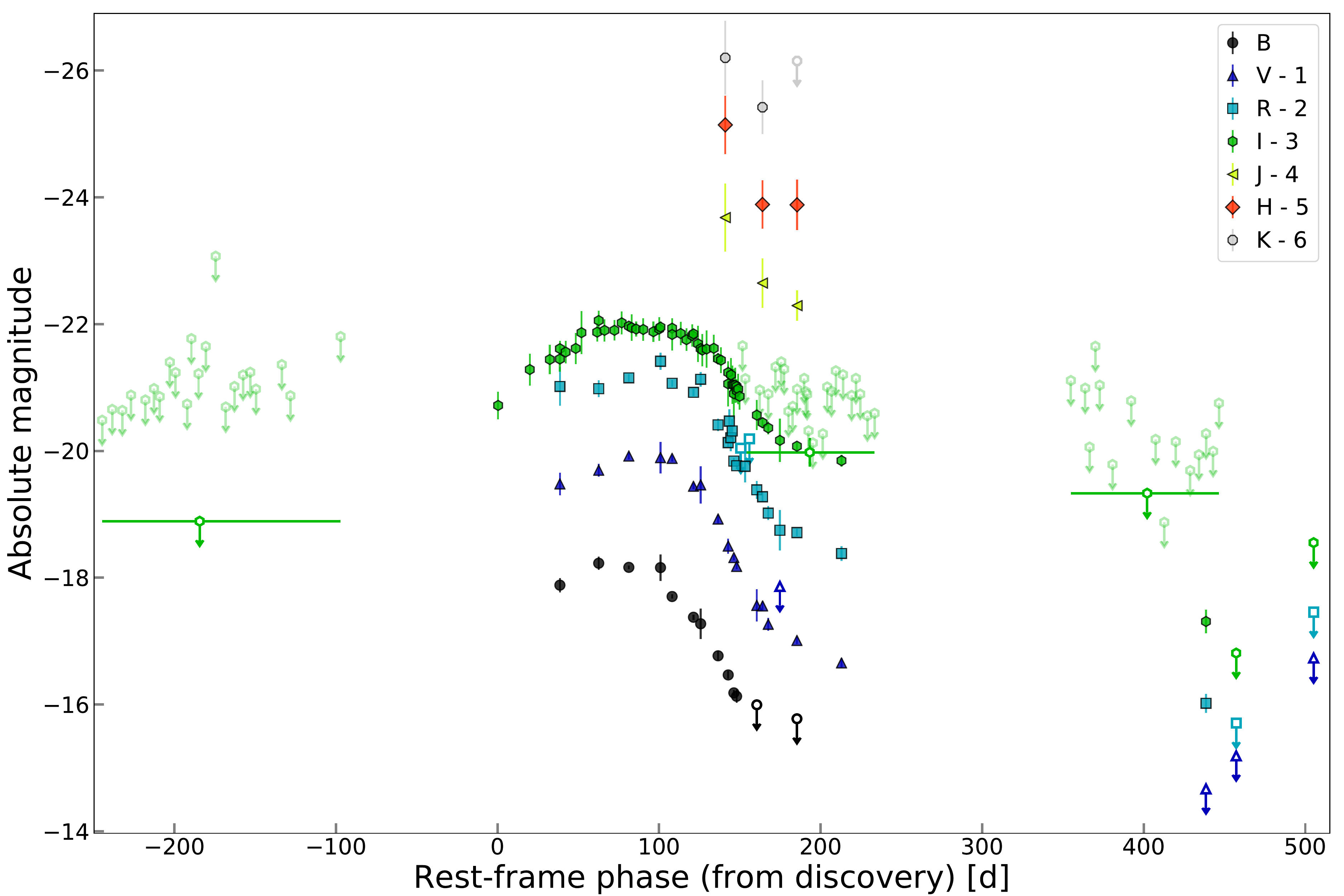}
\caption{\textbf{Multi-band photometric evolution of \sn{}} $-$ Upper-limits are indicated by an empty symbol with an arrow. Three groups of OGLE-IV images, where the SN was not detected, were stacked together in order to get three single deeper image. The upper-limits coming from each single image are still reported with shaded symbols. A phase error equal to the temporal range of the images staked has been attributed to the measurements inferred from the deeper images. As there is no constraint on the explosion epoch, the discovery epoch has been used as reference.}
\label{fig: snlc}
\end{figure}
\clearpage

\begin{figure}
\centering
\includegraphics[width=\textwidth]{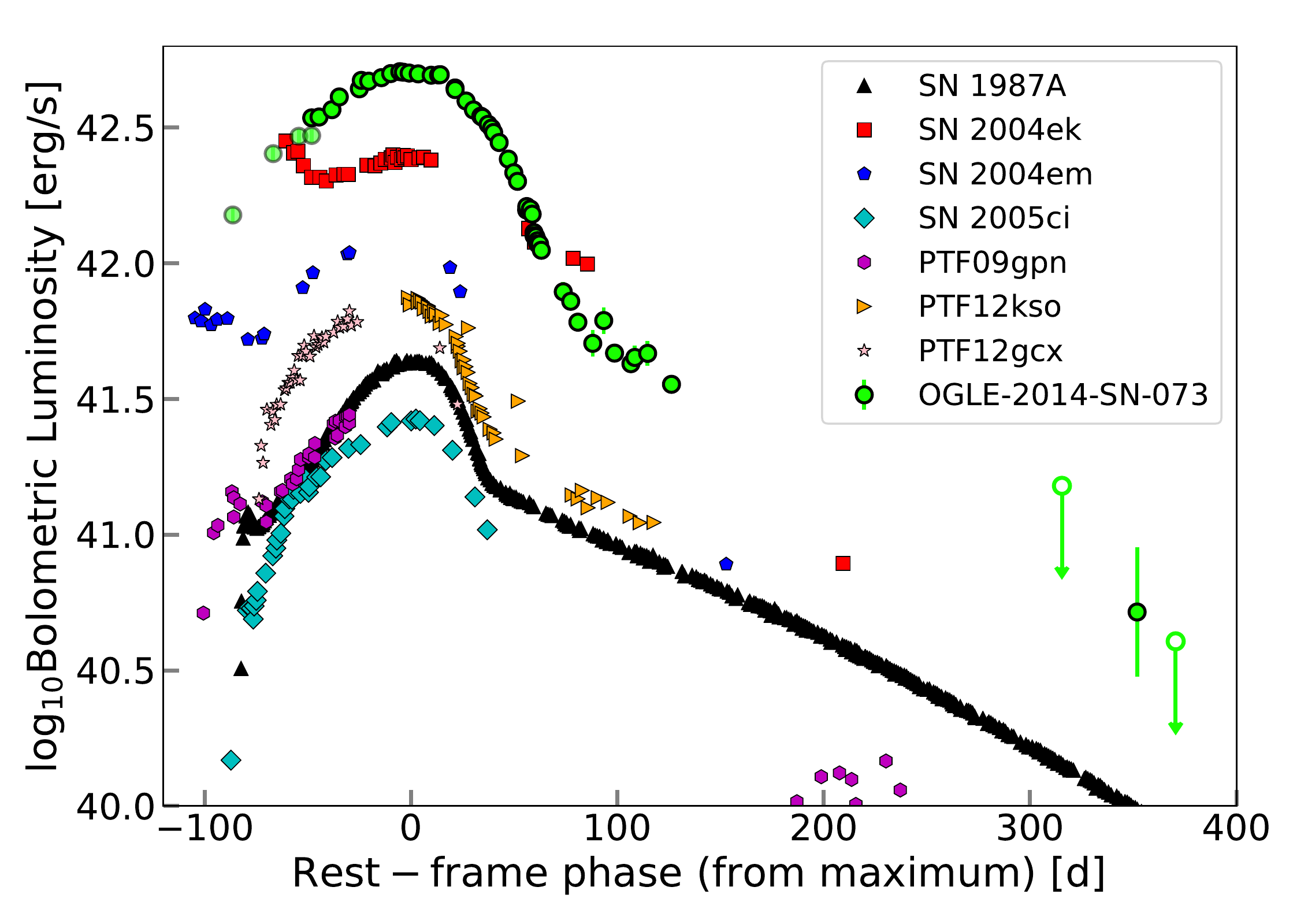}
\caption{\textbf{Comparison of the optical pseudo-bolometric lightcurve of \sn{} with a sample of the 1987A-like events} $-$ We considered the 1987A-like sample from \cite{Taddia2016}, i.e. SNe 2004ek, 2004em, 2005ci, PTF09gpn, PTF12kso and PTF12gcx. The phase is in rest frame and with respect maximum light.}
\label{fig: bol_87A}
\end{figure}
\clearpage

\begin{figure}
\centering
\includegraphics[height=15cm]{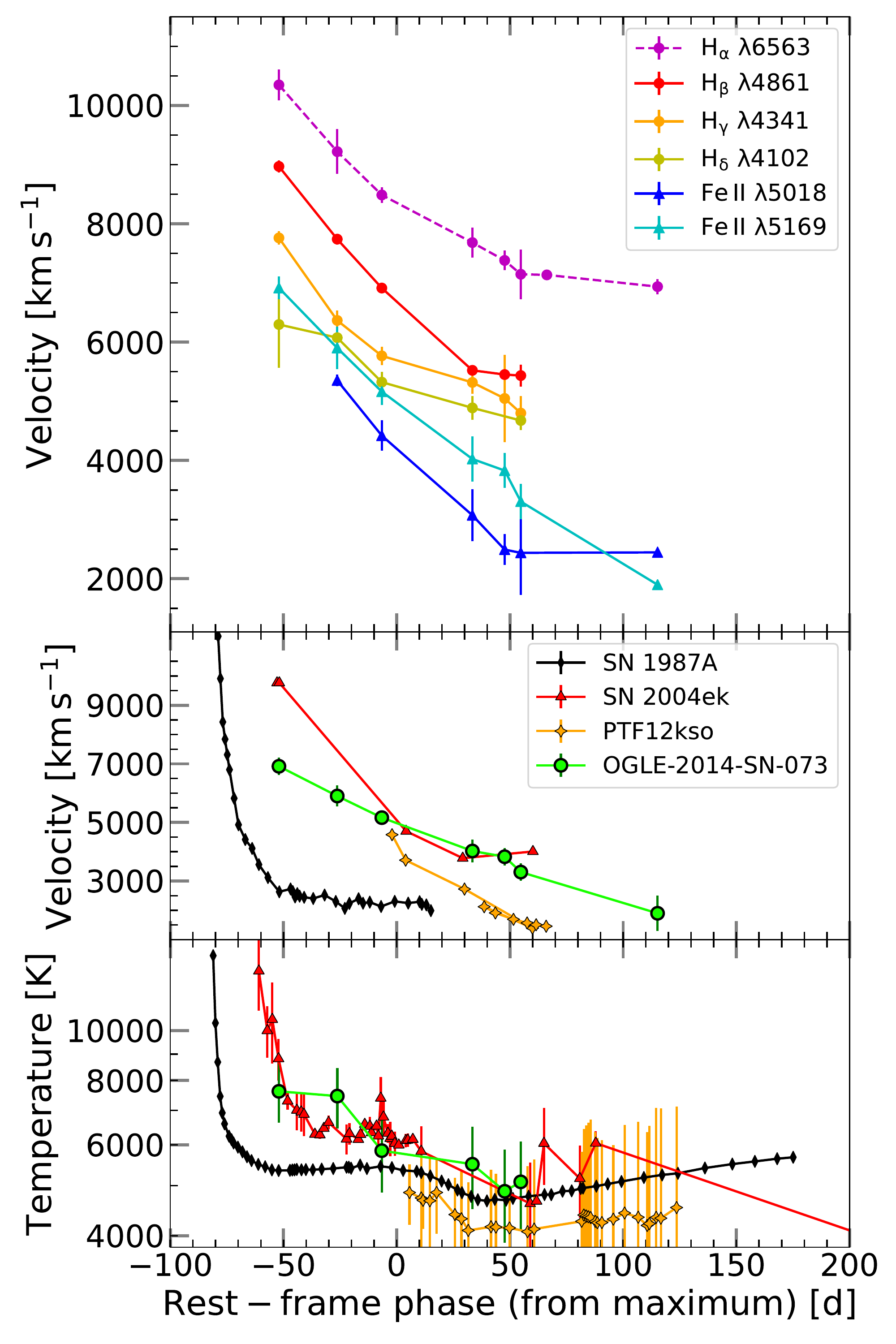}
\caption{\textbf{Line velocity and temperature evolution of \sn{}} $-$ \textbf{(a)}: Velocity evolution of the Balmer lines and the \ion{Fe}{ii} $\lambda$5018 and $\lambda$5169 of \sn{}. Velocities were measured from the Doppler-shift of the line, taken from the minimum of the absorption feature. \textbf{(b)}: Comparison of the \ion{Fe}{ii} $\lambda$5169 velocity of \sn{} with that of SN 1987A and the best covered 1987A-like events from the sample of \cite{Taddia2016}. \textbf{(c)}: temperature evolution of \sn{} and comparison with the SNe considered in the panel (b). Temperatures were measured by fitting the spectra with a black-body. The legend is the same as the above panel. Note that the scale is logarithmic.}
\label{fig: super}
\end{figure}
\clearpage

\vspace{-5cm}
\begin{figure}
\centering
\includegraphics[height=11cm]{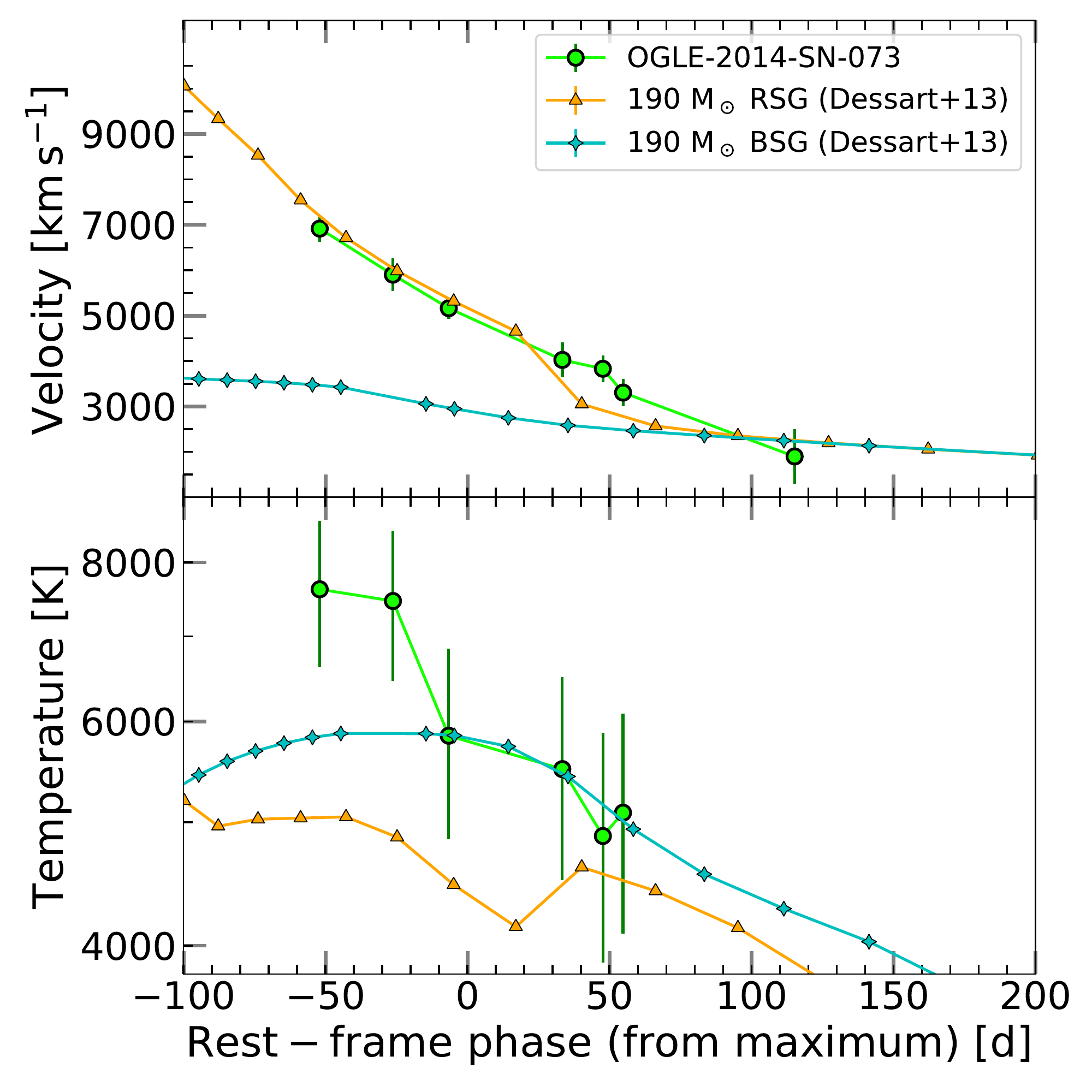}
\caption{\textbf{Comparison of the velocity and temperature evolution of \sn{} with PISN models} $-$ \textbf{(a)}: Comparison of the \ion{Fe}{ii} $\lambda$5169 velocity of \sn{} (green circles) with that of the PISN models from \cite{Dessart2013}, in particular the $M_{\rm{ZAMS}}=190$~\Msun{} progenitors exploding as a RSG (orange triangles) and BSG (cyan stars). \textbf{(b)}: temperature evolution of \sn{} and comparison with the above mention PISN models. The legend is the same as the above panel.
The velocities of \sn{} quantitatively match those of the RSG progenitor model, while the temperatures more closely resemble the BSG progenitor model. The disparity in the velocities between RSG and BSG comes from the fact that at maximum light in the latter model, the photosphere has receded to the slow moving He-core. In the former meanwhile, it is still in the faster moving partially-ionised H-rich envelope.}
\label{fig: super_2}
\end{figure}
\clearpage

\begin{table}
 \centering
 \begin{minipage}{140mm}
 \begin{tabular}{@{}lccc}
   \multicolumn{4}{c}{\textbf{Photometry}}\\
    Telescope & Instrument & FoV & Filters\\
    \hline
    NTT & EFOSC2 & $4.1'\times4.1'$ & \textit{BVRi}\\
     & SOFI & $4.92 '\times4.92 '$& \textit{JHK}\\
    1.3-m Warsaw telescope& 32-MOSAIC & $1.5~\rm{deg}^2$& \textit{I}\\
    LCO 1m-04 & Sinistro & $26.5'\times26.5'$ & \textit{grRiz}\\
    LCO 1m-09 & Sinistro & $26.5'\times26.5'$ & \textit{grRiz}\\
    VLT & FORS2 & $6.8'\times6.8'$ & \textit{VRI}\\
\hline
\\
\\
\end{tabular}
\bigskip
 \begin{tabular}{@{}lcccc}

  \multicolumn{5}{c}{\textbf{Spectroscopy}}\\
 Telescope & Instrument & Grating & Slit & Resolution [R]\\
  \hline
 NTT & EFOSC2 & Gr\#13 & 1.0$''$ & 355\\
Gemini South & GMOS & R400+G5325 & 1.5$''$ & 640\\
&&B600+G5323 & 1.5$''$ & 1250 \\
\end{tabular}

  \caption{Instrumental configurations used for the follow-up campaign of \sn{}.}
\label{tab: instr}
\end{minipage}
\end{table}
\clearpage

\end{document}